\documentclass[aip,jcp,amsmath,amssymb,reprint]{revtex4-2}

\usepackage{graphicx}
\usepackage{dcolumn}
\usepackage{bm}
\usepackage[utf8]{inputenc}
\usepackage[T1]{fontenc}
\usepackage{mathptmx}
\usepackage{natbib}
\usepackage{enumitem}   
\usepackage[table,xcdraw]{xcolor}
\usepackage{array}
\usepackage{lipsum}
\usepackage[ruled,vlined]{algorithm2e}
\SetKwInOut{KwInput}{Input}
\SetKwInOut{KwOutput}{Output}
\usepackage{capt-of}

\begin{document}

\title{Electronic Structure Calculations from Occupation Numbers on Quantum Computers}

\author{Edison X. Salazar}
 \email{edison.salazar@dipc.org}
\affiliation{Donostia International Physics Center (DIPC), 20018 Donostia, Euskadi, Spain.}

\author{Juan Felipe Huan Lew-Yee}
\email{felipe.lew.yee@quimica.unam.mx}
\affiliation{Departamento de F\'isica y Qu\'imica Te\'orica, Facultad de Qu\'imica, Universidad Nacional Aut\'onoma de M\'exico, M\'exico City, C.P. 04510, M\'exico}
\affiliation{Departamento de Matem\'aticas, Universidad Nacional Aut\'onoma de M\'exico, M\'exico City, C.P. 04510, M\'exico}

\author{Mario Piris}
\email{mario.piris@ehu.eus}
\affiliation{Donostia International Physics Center (DIPC), 20018 Donostia, Euskadi, Spain}
\affiliation{Euskal Herriko Unibertsitatea (EHU), 20018 Donostia, Euskadi, Spain}
\affiliation{Basque Foundation for Science (IKERBASQUE), 48009 Bilbao, Euskadi, Spain}

\date{\today}

\begin{abstract}
We present a quantum-classical algorithm for electronic structure calculations that dramatically reduces the quantum measurement cost of variational quantum eigensolver (VQE) approaches. While conventional VQE methods require measurements scaling as $\mathcal{O}(M^4)$ with system size $M$, the proposed occupation-number VQE (ON-VQE) reduces this cost to $\mathcal{O}(M/2)$ by avoiding reduced density matrix (RDM) measurements and relying exclusively on ONs. The method exploits only the diagonal elements of the one-particle RDM in the natural orbital representation, where occupations are obtained directly from computational-basis measurement outcomes. By restricting the variational ansatz to double excitations within orbital subspaces associated with electron pairs, the required measurements can be grouped into a small number of qubit-wise commuting observables, yielding an efficient and scalable measurement strategy. The approach is validated through simulations and executions on quantum hardware for the cubic H$_8$ cluster, demonstrating the feasibility of extracting accurate ONs from quantum measurements and evaluating electronic energies within the natural orbital functional (NOF) framework. Across representative molecular systems, the extracted ONs enable accurate energy evaluation with state-of-the-art NOFs while maintaining a dramatically reduced measurement cost. These results establish a scalable route toward quantum simulation of strongly correlated electronic systems, demonstrating that accurate electronic energies can be obtained from quantum measurements of ONs alone.
\end{abstract}


\maketitle

Electronic structure theory is undergoing a transformation driven by hybrid classical–quantum computer approaches. Among them, the Variational Quantum Eigensolver (VQE) has emerged as a leading framework for exploiting near-term quantum devices in chemistry and materials science \cite{Harville2026}. However, its practical applicability is severely limited by measurement overhead and circuit depth constraints in noisy intermediate-scale quantum (NISQ) hardware \cite{Cao2019}. In standard implementations, evaluating the ground-state energy requires measuring the electronic Hamiltonian, which involves a number of Pauli operators that scale as $\mathcal{O}(M^4)$ with the system size $M$, leading to prohibitive sampling costs. Although fault-tolerant quantum algorithms are expected to overcome these limitations, current devices remain constrained by noise and coherence times. 

As a result, the efficiency of quantum measurements becomes the dominant factor determining VQE performance. This bottleneck has motivated the development of alternative formulations based on reduced quantities, particularly reduced density matrices (RDMs) \cite{Smart2021,Garrett2026}, which offer a more compact representation of the quantum state while preserving the information required for energy evaluation. Among these approaches, the one-particle RDM (1RDM) has emerged as a particularly attractive quantity, drastically reducing the measurement cost relative to Hamiltonian-based approaches while improving robustness to measurement noise \cite{Tilly2021}. Its relevance is further supported by the existence of a universal functional \cite{Gilbert1975,Valone1980}. Although the exact functional is unknown, natural orbital functionals (NOFs) \cite{Piris2024,Piris2024a,Lew-Yee2026} provide accurate approximations \cite{Piris2011,Piris2017,Piris2021,Lew-Yee2023,Lew-Yee2024,Mitxelena2026}, particularly in correlated regimes \cite{Mitxelena2020a,Mitxelena2020b,Mitxelena2022,Mitxelena2024,Lew-Yee2025b,Lew-Yee2025c}.

In parallel, alternative hybrid strategies have explored reducing quantum resources by replacing a single variational state with structured subspace expansions built from shallow circuits. In these approaches, the time-independent Schrödinger equation is projected onto a subspace of the Hilbert space, leading to an eigenvalue problem determined by measurements carried out on a quantum device and solved on a classical computer \cite{Motta2024}. Physically motivated symmetries, such as seniority, can be exploited to construct orthogonal basis states and restrict the coupling structure of the Hamiltonian, effectively trading circuit depth for a more structured measurement problem \cite{Patel2026}. These methods redistribute the computational effort between quantum state preparation, measurement, and classical post-processing, offering an alternative route to improving the efficiency of hybrid quantum–classical algorithms.

In this context, we recently introduced the NOF-VQE framework \cite{Lew-Yee2025}, where the 1RDM measured on a quantum device is combined with classical NOF approximations to evaluate the energy, significantly reducing measurement costs while preserving accuracy. Despite these advances, reconstructing the full 1RDM still requires a number of measurements that scale as $\mathcal{O}(M^2)$ with system size, along with nontrivial classical post-processing such as basis transformations and diagonalization. This limitation raises a more fundamental question: whether the full 1RDM is necessary for accurate energy reconstruction within the NOF framework, or if a reduced set of quantities can capture the essential physics at a lower cost.

In the natural orbital (NO) representation, the 1RDM is diagonal by construction, enabling a measurement strategy based solely on occupation numbers (ONs). Given a quantum ground state $|\Psi_{gs}\rangle$, the ON of a spin-orbital $i$ reads
\begin{equation}
n_i = \langle \Psi_{gs} | \hat{a}_i^\dagger \hat{a}_i | \Psi_{gs} \rangle.
\end{equation}
Under the Jordan-Wigner transformation, the fermionic creation ($\hat{a}_i^\dagger$) and annihilation ($\hat{a}_i$) operators map onto Pauli operators as
\begin{equation}
\hat{a}_i^\dagger = \frac{X_i - iY_i}{2}\prod_{j<i} Z_j, \quad
\hat{a}_i = \frac{X_i + iY_i}{2}\prod_{j<i} Z_j.
\end{equation}
Substituting into the number operator $\hat{n}_i = \hat{a}_i^\dagger \hat{a}_i$, the Jordan--Wigner string cancels, yielding
\begin{equation}
\hat{n}_i = \frac{1}{2}(1 - Z_i).
\end{equation}
Thus, the ON is obtained directly from a single-qubit expectation value,
\begin{equation}
\label{eq:on_eva}
n_i = \frac{1}{2}\left(1 - \langle Z_i \rangle\right).
\end{equation}
This result shows that ONs can be obtained from local, single-qubit measurements of $Z_i$, avoiding the nonlocal operators involved in measuring the full 1RDM. Consequently, the measurement cost scales as $\mathcal{O}(M)$ with system size. Within the NOF framework, this enables the evaluation of the energy without explicit reconstruction of the 1RDM, using only ONs obtained from local measurements, provided that the NO basis is defined.

State-of-the-art NOFs are typically based on pairwise approximations \cite{Piris2018e}, such as PNOF5 \cite{Piris2011, Piris2013a, Piris2013c}, PNOF7 \cite{Piris2017, Mitxelena2018a, Piris2018b}, and GNOF \cite{Piris2021, Lew-Yee2025b}. These electron-pairing-based functionals have proven particularly effective for describing strongly correlated systems and offer significant advantages from both theoretical and practical perspectives \cite{Lew-Yee2023a,RiveroSantamaria2024,Piris2024b}. They further ensure identical ONs for spin-up and spin-down electrons ($n_{p} = n_{\bar{p}}$), regardless of the total spin. Exploiting this property, the measurement cost can be reduced to $\mathcal{O}(M/2)$, since it suffices to measure a single spin sector. In the following, we focus on singlet states; for multiplet states, see Ref.~\cite{Piris2019}.

A key aspect of the proposed approach is the choice of the variational ansatz. Unlike conventional VQE methods, in which the quantum circuit explicitly encodes the electronic wavefunction and directly yields the correlation energy, ON-VQE uses the NOF to steer how electronic correlation is handled on the quantum device, requiring only precise measurements of the ONs of the quantum state. As a result, the quantum circuit only needs to produce accurate ONs. Nevertheless, current hardware limitations favor shallow circuits, as increasing circuit depth leads to greater noise sensitivity and optimization challenges on NISQ devices. A suitable balance must therefore be achieved between circuit efficiency and the flexibility required to represent physically meaningful occupation distributions.

To address these challenges, we adopt an ansatz based on the generating wavefunction of PNOF5 \cite{Piris2013a}, given by
\begin{equation}
\left|\Psi_{gs}\right\rangle =\prod_{g=1}^{\mathrm{N}/{2}}\hat{\psi}_{g}^{\dagger}\left|0\right\rangle = 
\prod_{g=1}^{\mathrm{N}/{2}} [\sqrt{n_{g}}\hat{a}_{g}^{\dagger}\hat{a}_{\bar{g}}^{\dagger}-\sum_{p\in\Omega_{g}^{\prime}}\sqrt{n_{p}}\hat{a}_{p}^{\dagger}\hat{a}_{\bar{p}}^{\dagger}]\left|0\right\rangle
\label{wf1}
\end{equation}
where $\hat{\psi}_{g}^{\dagger}$ is a geminal creation operator that generates an electron pair with opposite spins \cite{Surjan1999}. This construction partitions the orbital space $\Omega$ into $\mathrm{N}/2$ subspaces $\Omega_g$, each containing one orbital $\left|g\right\rangle$ ($g \leq \mathrm{N}/2$) and $\mathrm{N}_g$ orbitals $\left|p\right\rangle$ ($p > \mathrm{N}/2$). The subset $\Omega_{g}^{\prime} = \Omega_g \setminus {g}$ includes only orbitals above the $\mathrm{N}/2$ level. The expansion coefficients of each geminal are determined by the ONs of the corresponding subspace, which sum to unity, ensuring that the trace of the 1RDM equals the total number of electrons, $\mathrm{N}$. In general, $\mathrm{N}_g$ may vary across subspaces; here it is fixed for all $\Omega_g$ and set to its maximum value, determined by the number of basis functions (qubits).

The ansatz (\ref{wf1}) can be rewritten as
\begin{equation}
\begin{array}{c}
\left|\Psi_{gs}\right\rangle ={\displaystyle \prod_{g=1}^{\mathrm{N}/2}}\sqrt{n_{g}}\left(1-\hat{T_{g}}\right)\hat{a}_{g}^{\dagger}\hat{a}_{\bar{g}}^{\dagger} \left|0\right\rangle 
\end{array}\label{CC}
\end{equation}
\noindent where
\begin{equation}
\hat{T_{g}}=\frac{1}{\sqrt{n_{g}}}\left({\displaystyle \sum_{p\in\Omega_{g}^{\prime}}}\sqrt{n_{p}}\hat{a}_{p}^{\dagger}\hat{a}_{\bar{p}}^{\dagger}\right)\hat{a}_{\bar{g}}\hat{a}_{g}.
\end{equation}

To enable implementation on a quantum computer, the non-unitary structure in Eq.~(\ref{CC}) is mapped onto a unitary form using exponentials of anti-Hermitian pair excitation operators. Defining $\hat{\tau}_g^p = \hat{a}_p^{\dagger}\hat{a}_{\bar{p}}^{\dagger}\hat{a}_{\bar{g}}\hat{a}_g$, the ansatz becomes
\begin{equation}
\hat{U}(\boldsymbol{\theta}) = \prod_{g=1}^{\mathrm{N}/2}
\exp\left[\sum_{p\in\Omega_g'} \theta_g^p \left(\hat{\tau}_g^p - \hat{\tau}_g^{p\dagger}\right)\right].
\end{equation}
This construction preserves particle number and seniority while defining a pair-correlated ansatz analogous to a seniority-zero (pair-coupled cluster) expansion.

\begin{figure*}[t]
\includegraphics[scale=0.25]{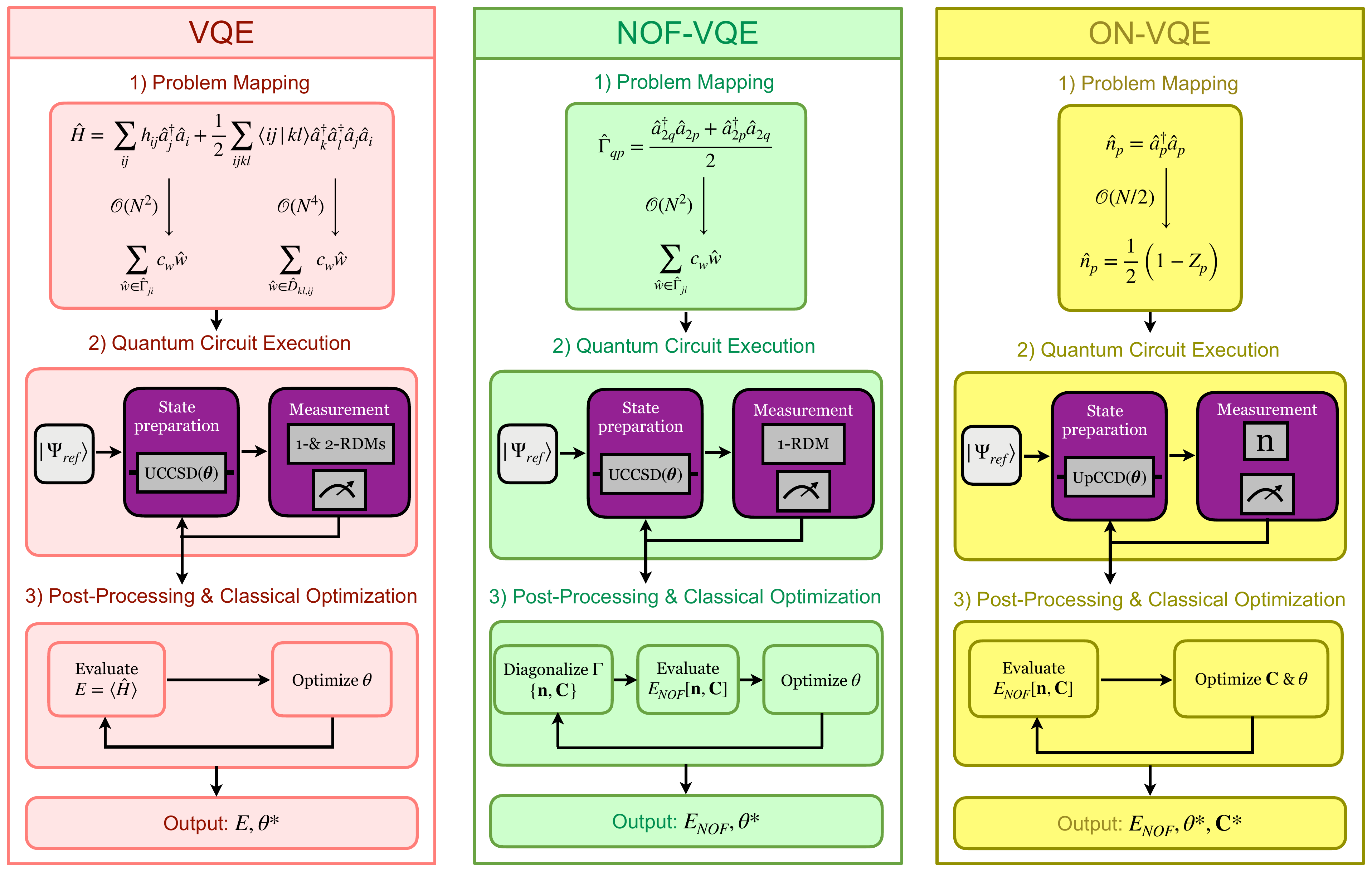}
\caption{\label{fig:workflow_v2} Comparison of the computational workflows employed in VQE, NOF-VQE, and ON-VQE. Here, $\mathbf{C}$ denotes the coefficient matrix whose columns define the NOs in the underlying atomic orbital basis.}
\end{figure*}

In practice, the exponential is approximated using a first-order Trotter decomposition, yielding a product of exponentials of individual pair excitation operators that can be mapped onto parameterized two-body gates. Each factor corresponds to a single pair-excitation unitary and is implemented as an individual parameterized gate in the quantum circuit. This implementation retains the pair-correlated structure and is compatible with standard variational quantum algorithms.

The parameters $\{\theta_g^p\}$ are treated variationally and optimized on the quantum processor to minimize a NOF-based energy evaluated from the ONs of the quantum state. The circuit can be implemented in a quantum computing framework with native support for fermionic pair-double excitations, where each Trotterized term is mapped onto a parameterized two-body gate.

The formulation is expressed in the NO representation. Since these orbitals are not known a priori, they are determined self-consistently within a hybrid scheme in which the orbital basis is optimized classically, while the quantum processor is used to optimize the circuit parameters. In practice, we employ the adaptive momentum (ADAM) algorithm for orbital optimization \cite{Lew-Yee2025b} and the sequential least squares programming (SLSQP) method for quantum variational parameters. The resulting orbital basis accelerates convergence and provides a compact representation. When a reliable approximation to the NO basis is available, the procedure is reduced to the optimization of circuit parameters only.

Restricting the ansatz to pair-double excitations within each subspace leads to a compact circuit representation that preserves the structure of the underlying wavefunction while remaining compatible with current NISQ hardware. Moreover, the ONs are obtained from observables that are diagonal in the computational basis and therefore belong to the same qubit-wise commuting (QWC) measurement group. Consequently, all occupations can be extracted simultaneously from a single set of circuit executions, eliminating the need for explicit 1RDM reconstruction. This constitutes a central advantage of ON-VQE over previous NOF-VQE, substantially reducing the quantum measurement overhead while preserving all the information required for NOF energy evaluation. Unlike conventional VQE and RDM-based approaches, ON-VQE extracts only the ONs from the variational quantum state, while the electronic energy is evaluated through the NOF. As a result, neither the Hamiltonian expectation value nor the full 1RDM needs to be reconstructed from quantum measurements.

Fig.~\ref{fig:workflow_v2} summarizes the workflow of the conventional VQE, NOF-VQE, and ON-VQE methods. While VQE requires the measurement of both RDMs to reconstruct the Hamiltonian expectation value, NOF-VQE reduces the measurement cost by targeting only the 1RDM. ON-VQE further simplifies the quantum workflow by directly measuring local ON operators, allowing the NOF energy to be evaluated without explicit 1RDM reconstruction. In contrast to VQE and NOF-VQE, the ON-VQE framework also performs a self-consistent optimization of the NOs, enabling the use of compact pair-based ansätze such as UpCCD in the NO representation. A reference implementation is available through our open-source GitHub repository \cite{Salazar2026ONVQE}.

The performance of the proposed framework is assessed in the following section through molecular benchmarks, quantum simulations, and hardware executions for the cubic H$_8$ cluster.

\subsection{Molecular Systems Benchmark}

\begin{figure}[htb]
\includegraphics[scale=0.22]{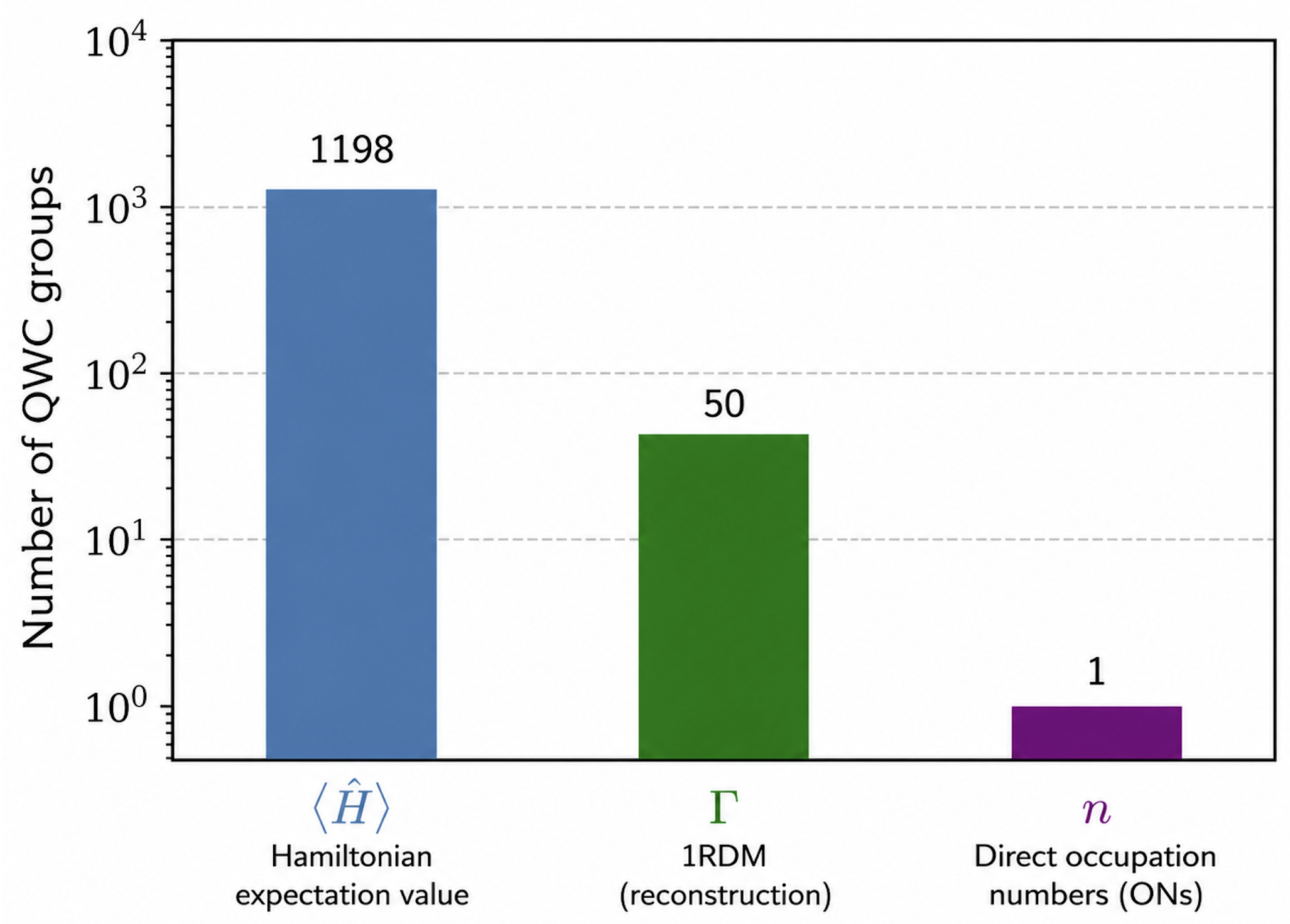}
\caption{\label{fig:repvalues} Representative values obtained for the largest molecular systems considered.}
\end{figure}

Fig.~\ref{fig:repvalues} compares the measurement overhead associated with three quantum evaluation strategies for electronic structure calculations: conventional VQE based on direct Hamiltonian expectation value measurements $\langle \hat H \rangle$, NOF-VQE based on the reconstruction of the 1RDM, and the present ON-VQE approach based on direct ON measurements. The reported values are representative of the molecular systems considered in this work, namely H$_2$, LiH, Li$_2$, OH$^{-}$, FH, NeH$^{+}$, and F$_2$ in the STO-3G basis, and correspond to the largest measurement overhead observed within this set. The results reveal a clear hierarchy in measurement complexity, requiring 1198 QWC groups for conventional VQE, 50 groups for NOF-VQE, and only a single QWC group for ON-VQE.

This substantial reduction originates from the fact that the NOF energy depends exclusively on ONs in the NO representation. Since all ONs are extracted from measurement outcomes in the computational basis, the complete set of occupations can be obtained simultaneously from the same collection of bitstrings, eliminating the need for explicit reconstruction of the 1RDM and the measurement of off-diagonal RDM elements. Compared with the original NOF-VQE formulation based on the full 1RDM, ON-VQE exploits the diagonal structure of the 1RDM in the NO basis, such that all quantities required for energy evaluation belong to the same QWC measurement group and can therefore be obtained from a single measurement setting.

Beyond the reduction in measurement complexity, this simplification also decreases the quantum execution cost. Since each QWC group requires an independent set of circuit executions, reducing the number of measurement groups directly lowers the overall quantum workload during the variational optimization. For current NISQ devices, the resulting savings can be redirected toward increasing the number of shots per measurement, thereby improving statistical precision and enhancing robustness against noise.

\begin{table}[htb]
\centering
\caption{\label{tab:energies} Total energies (Ha) obtained from the ONs extracted with ON-VQE and evaluated using different NOFs at the equilibrium distance.}
\bigskip
\begin{ruledtabular}
\begin{tabular}{lcccccc}
Molecule & Qubits & $\langle \hat H \rangle$ & PNOF5 & PNOF7 & GNOF \\
\hline
H$_2$     & 4  &   -1.137 &    -1.137 &   -1.137 &   -1.137 \\
LiH       & 12 &   -7.882 &    -7.882 &   -7.882 &   -7.885 \\
Li$_2$    & 20 &  -14.664 &   -14.658 &  -14.659 &  -14.664 \\
OH$^-$    & 12 &  -74.080 &   -74.076 &  -74.076 &  -74.076 \\
FH        & 12 &  -98.596 &   -98.595 &  -98.595 &  -98.595 \\
NeH$^+$   & 12 & -126.811 &  -126.809 & -126.809 & -126.809 \\
F$_2$     & 20 & -196.050 &  -196.045 & -196.045 & -196.045 \\
\end{tabular}
\end{ruledtabular}
\end{table}

Table~\ref{tab:energies} reports the total energies obtained from the ONs generated by ON-VQE and evaluated using different NOFs. The expectation value $\langle \hat H \rangle$ corresponds to the conventional VQE algorithm and serves as the reference value for the present ansatz. The remaining columns report the energies obtained by evaluating different NOF approximations using the same set of ONs extracted from the quantum state. All the energies were computed using the lightning-exact simulator \cite{Asadi2024}. The purpose of this benchmark is not to assess the relative performance of different NOFs, but to demonstrate that the same set of ONs can be combined with multiple NOF approximations without additional quantum measurements.

For all systems considered, the energies obtained with PNOF5, PNOF7, and GNOF remain very close to the variational reference. This behavior is primarily due to the minimal STO-3G basis employed in the calculations, which leads to relatively small electron-pair subspaces ($N_g\leq2$). Under these conditions, the different NOF approximations naturally yield similar energies. More importantly, Table I demonstrates that a single set of ONs extracted from the quantum circuit is sufficient to evaluate different NOF approximations while requiring substantially fewer quantum measurements than previous RDM-based formulations. Consequently, improvements in the classical NOF approximation can be incorporated without modifying the quantum workflow or increasing the measurement cost. This separation between quantum data acquisition and classical energy evaluation makes ON-VQE a flexible framework for the continued development of NOF-based quantum algorithms.

\subsection{H$_8$ Benchmark}

\begin{figure}[ht]
\includegraphics[width=\columnwidth]{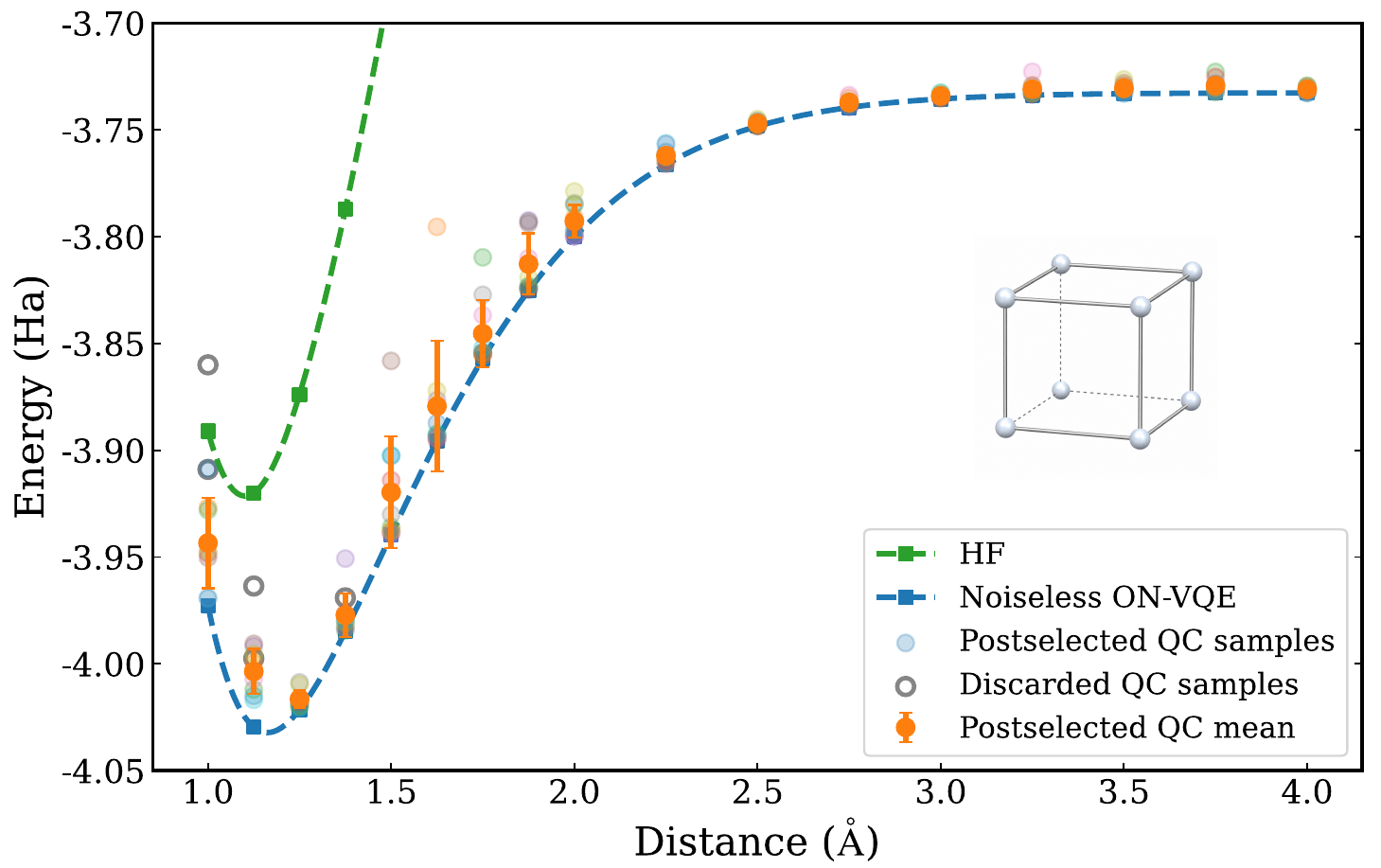}
\caption{\label{fig:pes_h8_hybrid} Potential energy curve of the cubic H$_8$ cluster obtained from ONs measured on quantum hardware.}
\end{figure}

We assessed the performance of ON-VQE for the symmetric dissociation of the cubic H$_8$ cluster in the STO-3G basis. In this approach, ONs are extracted directly from measurements in the NO representation and used to evaluate the NOF energy. Figure~\ref{fig:pes_h8_hybrid} compares the Hartree--Fock (HF) reference, the noiseless ON-VQE potential energy curve (PEC), the individual post-selected measurements obtained on quantum hardware, and their corresponding post-selected averages. Quantum hardware calculations were performed on the IBM BasqueCountry processor, while the circuit parameters were optimized using the lightning-exact simulator. Quantum executions used IBM Runtime optimization level 3, resilience level 2, and 10\,000 shots per circuit evaluation. Each point on the PEC corresponds to the average of the post-selected hardware executions. Individual hardware executions are also shown, with discarded samples indicating measurements that violate the physical ON constraints of the electron-pair model owing to hardware errors. The complete error-mitigation protocol is described below. 

The post-selected quantum results closely follow the noiseless ON-VQE reference throughout the entire dissociation process. Despite hardware noise, the averaged quantum measurements accurately reproduce both the equilibrium region and the dissociation limit. While individual quantum samples exhibit a larger dispersion near equilibrium, the spread decreases as the system approaches dissociation, where the dominant ONs naturally converge toward fractional values close to one-half. In this regime, measurement errors produce smaller perturbations in the NOF energy, leading to excellent agreement between quantum hardware and noiseless simulations.

A key advantage of ON-VQE is that the measured observables correspond directly to physically meaningful ONs satisfying well-defined N-representability conditions. This enables a simple, physics-informed error-mitigation strategy operating directly in ON space, without reconstructing RDMs. After measurement, the occupations are projected onto the physically allowed interval,
\begin{equation}
0 \leq n_i \leq 1,
\end{equation}
while enforcing the electron-pair constraint
\begin{equation}
n_g + \sum_i n_{p_i} = 1,
\end{equation}
within each subspace $\Omega_g$. A polarization-recovery step is subsequently applied to compensate for the tendency of hardware noise to drive occupations toward the center of their allowed interval. Finally, corrections are accepted only when consistent with the statistical uncertainty estimated from finite-shot sampling, while samples requiring excessive corrections are discarded during post-selection. Further implementation details of the error-mitigation protocol are provided in the Supporting Information.

Beyond improving robustness against noise, the ON-VQE framework further reduces the measurement cost. In the present STO-3G H$_8$ calculation, each electron-pair subspace contains one strongly occupied and one weakly occupied orbital, so that $ n_p = 1 - n_g $. Consequently, only one occupation per pair must be measured explicitly, halving the number of required observables without introducing additional approximations. This reduction follows directly from the pair structure of the NOF formalism and becomes increasingly advantageous for larger systems.

It is important to emphasize that, unlike classical NOF implementations such as DoNOF/PyNOF, the occupations in noiseless ON-VQE are not generated through constrained parametrizations. In classical NOF methods, N-representability is enforced through explicit parametrizations of the ONs. In contrast, ON-VQE occupations arise directly from a quantum state and therefore satisfy the representability conditions automatically in the noiseless limit. The restoration procedure becomes necessary only in hardware implementations, where noise perturbs the measured occupations away from the physical manifold.

These results highlight an important advantage of ONs as quantum observables for electronic-structure calculations. Their direct physical interpretation enables physics-informed error-mitigation and post-processing strategies that are difficult to implement in conventional energy-based VQE formulations. Although the development of advanced mitigation techniques lies beyond the scope of the present work, the occupation-space purification protocol introduced here already yields a substantial improvement in hardware performance and suggests promising directions for future research.

In summary, we have introduced ON-VQE, a quantum-classical algorithm that directly extracts ONs in the NO representation, eliminating the need for explicit reconstruction of the 1RDM. By exploiting the fact that the NOF energy depends only on ONs, the proposed approach reduces the quantum measurement overhead to a minimal set of observables while retaining the information required for accurate energy evaluation. The results obtained for representative molecular systems and for the cubic H$_8$ cluster on both noiseless simulations and quantum hardware demonstrate that direct ON measurements provide a viable alternative to RDM-based quantum workflows. By reducing the quantum information required for energy evaluation to ON measurements alone, ON-VQE significantly alleviates one of the main bottlenecks of NOF-VQE and related VQE-based approaches, providing a scalable framework for quantum electronic-structure calculations on near-term quantum devices. More generally, ON-VQE demonstrates that physically motivated reduced quantities can provide an effective interface between quantum state preparation and classical electronic-structure theory, opening new opportunities for measurement-efficient quantum algorithms.

\textit{Data availability.} Data supporting this study are available in the article and its Supplementary Material. Additional data are available from the corresponding author upon reasonable request.

\textit{Supplementary Information.} 
Supplementary Information contains implementation details, including the validation of the noiseless ON-VQE implementation, the self-consistent algorithm, the occupation-space error-mitigation protocol, and quantum hardware settings.

\textit{Acknowledgments}. Financial support comes from the Eusko Jaurlaritza (Basque Government), Ref.: IT2067-26. E. X. Salazar acknowledges the DIPC and the IKUR program under reference 2025/25 for post-doctoral funding. The authors acknowledge the technical and human support provided by both the DIPC Supercomputing Center and IZO-SGI (SGIker) of EHU and European funding (ERDF and ESF).

\renewcommand{\thefigure}{S\arabic{figure}}
\renewcommand{\arraystretch}{1.2}
\setcounter{figure}{0}

\renewcommand{\thetable}{S\Roman{table}}
\setcounter{table}{0}

\section*{Supplemental Material}

\section{Validation of the noiseless ON-VQE implementation}
\label{sec:validation}

Figure~\ref{fig:pes_h8} compares the reference PNOF7 potential energy curve (PEC) obtained with the classical PyNOF implementation and the corresponding noiseless ON-VQE results. The excellent agreement between both curves demonstrates that, in the absence of hardware noise, the quantum implementation faithfully reproduces the classical NOF energy throughout the dissociation process.

\begin{center}
\includegraphics[scale=0.55]{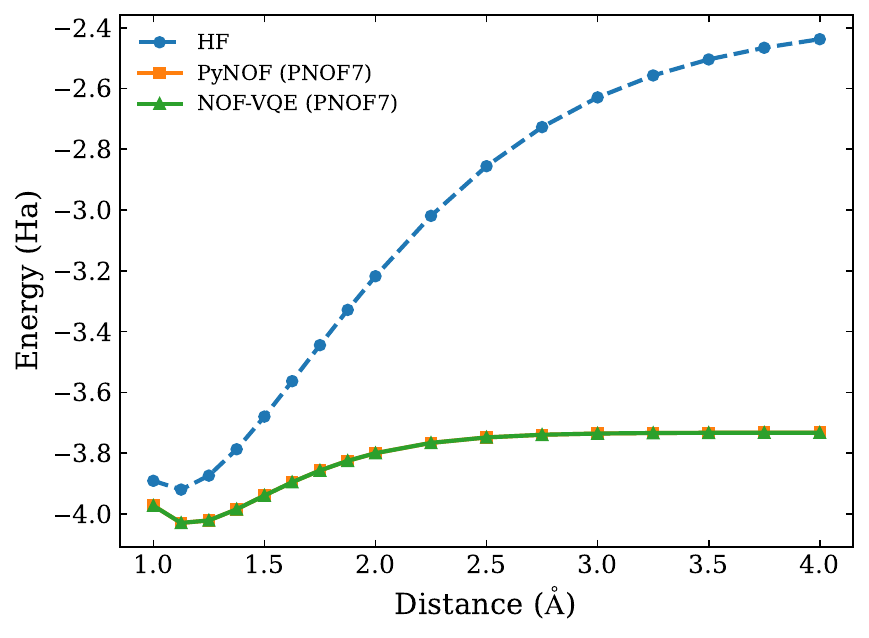}
\captionof{figure}{Comparison between the classical PyNOF and noiseless ON-VQE PECs for the symmetric dissociation of the cubic H$_8$ cluster using PNOF7.}
\label{fig:pes_h8}
\end{center}

\section{Self-consistent ON-VQE algorithm}
\label{sec:SI_on-vqe_algoritm}
\begin{algorithm}[H]
\DontPrintSemicolon

\KwInput{
Molecular geometry, basis set,
initial molecular orbitals,
initial circuit parameters $\boldsymbol{\theta}$
}

\KwOutput{
Optimized energy $E$,
ONs $\{n_p\}$,
NOs $\{C\}$
}

Initialize molecular orbitals\;

Construct the UpCCD ansatz\;

\While{orbital optimization not converged}{

Build molecular integrals in the current NO basis\;

\While{circuit optimization not converged}{

Prepare the variational state
$|\Psi(\boldsymbol{\theta})\rangle$\;

Measure $\langle Z_p\rangle$\;

Compute
$
n_p=\frac12(1-\langle Z_p\rangle)
$\;

Infer complementary occupations from the
electron-pair constraints (when applicable)\;

Apply occupation-space error mitigation
(hardware only)\;

Evaluate the NOF energy\;

Update $\boldsymbol{\theta}$ using SLSQP\;

}

Update the NOs using ADAM\;

}

\Return{
$E,\{n_p\},\{C\}$
}

\caption{Self-consistent ON-VQE workflow.}

\label{alg:scnofvqe}

\end{algorithm}

\subsection*{Optimization protocol}

The ON-VQE calculations are performed through a nested
self-consistent optimization of the NOs and
the quantum circuit parameters. For the first geometry, the molecular orbitals are initialized from the Hartree--Fock solution, while the variational parameters are initialized to 0.1.

For PEC calculations, the optimized NOs and variational parameters obtained at one geometry are used as the initial guess for the next geometry. This warm-start strategy substantially accelerates convergence along the dissociation path while preserving the self-consistent optimization procedure.

\section{Physics-informed ON error mitigation}
\label{sec:SI_error_mitigation}

Since ON-VQE directly measures ONs, error mitigation can be performed directly in ON space rather than after reconstructing RDMs.

Because the measured ONs are the variables entering the NOF energy functional, the correction procedure is designed to preserve the physical constraints imposed by the electron-pair model while remaining consistent with the statistical uncertainty arising from finite-shot sampling.

The complete mitigation protocol consists of three consecutive stages.

\subsection{Physical projection}

The measured ONs are first projected onto the physically admissible NOF manifold. For each electron-pair subspace $\Omega_g$, the following conditions are enforced

\begin{equation}
\nonumber 0 \le n_i \le 1 \;,\;\; n_g+\sum_{p\in\Omega_g}n_p =1 \;,\;\; n_g \ge \sum_{p\in\Omega_g}n_p ,
\end{equation}
which guarantee the physical consistency of the measured ONs.

\subsection{Polarization recovery}

After the physical projection, hardware noise tends to reduce the occupation polarization by driving the ONs toward one-half.

A mild polarization restoration is therefore applied by rescaling the pair polarization
\begin{equation}
\nonumber P_g=n_g-\sum_{p\in\Omega_g}n_p,
\end{equation}
while preserving the normalization of each electron pair. The recovery factor is selected adaptively according to the average polarization of the measured state, preventing overcorrection in the strongly correlated regime.

\subsection{Trust-region validation and post-selection}

The polarization recovery is applied only when statistically justified. A trust region is constructed from the expected ON uncertainty arising from finite-shot sampling,
\begin{equation}
\nonumber \sigma_{n_i} = \frac{\sqrt{1-\langle Z_i\rangle^2}}{2\sqrt{N_{\rm shots}}},
\bigskip
\end{equation}
which defines an admissible correction radius in ON space. When the proposed correction exceeds this statistical region, only the mandatory physical projection is retained.

Finally, corrected configurations are subjected to a statistical consistency test based on pair normalization errors, occupation corrections, and branch-condition violations. Configurations satisfying these criteria are accepted for the statistical averages reported in the main text, whereas inconsistent measurements are discarded.

\subsection{Quantum hardware settings}

All hardware calculations reported in this work were performed on the IBM BasqueCountry quantum processor. Quantum executions employed IBM Runtime optimization level 3, resilience level 2, and 10\,000 measurement shots per circuit evaluation.

Each point of the potential energy curve corresponds to the average of ten independent hardware executions after
application of the occupation-space error-mitigation protocol described above.

The parameters reported in Table~\ref{tab:mitigation_parameters} were found to provide a robust compromise between statistical filtering, ON recovery, and preservation of the physical constraints throughout all H$_8$ hardware calculations reported in this work.

\begin{center}
\captionof{table}{
Parameters employed in the occupation-space error-mitigation
protocol used for the hardware calculations.
}
\label{tab:mitigation_parameters}

\begin{tabular}{lc}
\hline
Parameter & Value \\
\hline
IBM Runtime optimization level & 3 \\
IBM Runtime resilience level & 2 \\
Shots per circuit & 10\,000 \\
$k_\sigma$ (trust region) & 3.5 \\
Pair-sum tolerance & 0.05 \\
Maximum ON correction & 0.10 \\
Maximum pair correction & 0.10 \\
Strong-correlation threshold & $\overline{P}<0.25$ \\
Intermediate polarization & $0.20<\overline{P}\le0.60$ \\
High polarization & $\overline{P}>0.60$ \\
$\lambda_{\rm pol}$ & 1.00, 1.05, or 1.10 \\
\hline
\end{tabular}
\end{center}


\begin{thebibliography}{37}
\makeatletter
\providecommand \@ifxundefined [1]{%
 \@ifx{#1\undefined}
}%
\providecommand \@ifnum [1]{%
 \ifnum #1\expandafter \@firstoftwo
 \else \expandafter \@secondoftwo
 \fi
}%
\providecommand \@ifx [1]{%
 \ifx #1\expandafter \@firstoftwo
 \else \expandafter \@secondoftwo
 \fi
}%
\providecommand \natexlab [1]{#1}%
\providecommand \enquote  [1]{``#1''}%
\providecommand \bibnamefont  [1]{#1}%
\providecommand \bibfnamefont [1]{#1}%
\providecommand \citenamefont [1]{#1}%
\providecommand \@href[1]{\@@startlink{#1}\@@href}%
\providecommand \@@href[1]{\endgroup#1\@@endlink}%
\providecommand \@sanitize@url [0]{\catcode `\\12\catcode `\$12\catcode `\&12\catcode `\#12\catcode `\^12\catcode `\_12\catcode `\%12\relax}%
\providecommand \@@startlink[1]{}%
\providecommand \@@endlink[0]{}%
\providecommand \@url [1]{\endgroup\@href {#1}{\urlprefix }}%
\providecommand \urlprefix  [0]{URL }%
\providecommand \Eprint [0]{ }%
\providecommand \doibase [0]{https://doi.org/}%
\providecommand \selectlanguage [0]{\@gobble}%
\providecommand \bibinfo  [0]{\@secondoftwo}%
\providecommand \bibfield  [0]{\@secondoftwo}%
\providecommand \translation [1]{[#1]}%
\providecommand \BibitemOpen [0]{}%
\providecommand \bibitemStop [0]{}%
\providecommand \bibitemNoStop [0]{.\EOS\space}%
\providecommand \EOS [0]{\spacefactor3000\relax}%
\providecommand \BibitemShut  [1]{\csname bibitem#1\endcsname}%
\let\auto@bib@innerbib\@empty
\bibitem [{\citenamefont {Harville}\ \emph {et~al.}(2026)\citenamefont {Harville}, \citenamefont {Khurana}, \citenamefont {Grizzi},\ and\ \citenamefont {Liu}}]{Harville2026}%
  \BibitemOpen
  \bibfield  {author} {\bibinfo {author} {\bibfnamefont {T.}~\bibnamefont {Harville}}, \bibinfo {author} {\bibfnamefont {R.}~\bibnamefont {Khurana}}, \bibinfo {author} {\bibfnamefont {V.~F.}\ \bibnamefont {Grizzi}},\ and\ \bibinfo {author} {\bibfnamefont {C.}~\bibnamefont {Liu}},\ }\bibfield  {title} {\bibinfo {title} {{Recent Developments in VQE: Survey and Benchmarking}},\ } {\bibfield  {journal} {\bibinfo  {journal} {arXiv: 2602.11384}\ } (\bibinfo {year} {2026})}\BibitemShut {NoStop}%
\bibitem [{\citenamefont {Cao}\ \emph {et~al.}(2019)\citenamefont {Cao}, \citenamefont {Romero}, \citenamefont {Olson}, \citenamefont {Degroote}, \citenamefont {Johnson}, \citenamefont {Kieferov{\'{a}}}, \citenamefont {Kivlichan}, \citenamefont {Menke}, \citenamefont {Peropadre}, \citenamefont {Sawaya}, \citenamefont {Sim}, \citenamefont {Veis},\ and\ \citenamefont {Aspuru-Guzik}}]{Cao2019}%
  \BibitemOpen
  \bibfield  {author} {\bibinfo {author} {\bibfnamefont {Y.}~\bibnamefont {Cao}}, \bibinfo {author} {\bibfnamefont {J.}~\bibnamefont {Romero}}, \bibinfo {author} {\bibfnamefont {J.~P.}\ \bibnamefont {Olson}}, \bibinfo {author} {\bibfnamefont {M.}~\bibnamefont {Degroote}}, \bibinfo {author} {\bibfnamefont {P.~D.}\ \bibnamefont {Johnson}}, \bibinfo {author} {\bibfnamefont {M.}~\bibnamefont {Kieferov{\'{a}}}}, \bibinfo {author} {\bibfnamefont {I.~D.}\ \bibnamefont {Kivlichan}}, \bibinfo {author} {\bibfnamefont {T.}~\bibnamefont {Menke}}, \bibinfo {author} {\bibfnamefont {B.}~\bibnamefont {Peropadre}}, \bibinfo {author} {\bibfnamefont {N.~P.~D.}\ \bibnamefont {Sawaya}}, \bibinfo {author} {\bibfnamefont {S.}~\bibnamefont {Sim}}, \bibinfo {author} {\bibfnamefont {L.}~\bibnamefont {Veis}},\ and\ \bibinfo {author} {\bibfnamefont {A.}~\bibnamefont {Aspuru-Guzik}},\ }\bibfield  {title} {\bibinfo {title} {{Quantum Chemistry in the Age of Quantum Computing}},\ }\href {https://doi.org/10.1021/acs.chemrev.8b00803}
  {\bibfield  {journal} {\bibinfo  {journal} {Chemical Reviews}\ }\textbf {\bibinfo {volume} {119}},\ \bibinfo {pages} {10856} (\bibinfo {year} {2019})}\BibitemShut {NoStop}%
\bibitem [{\citenamefont {Smart}\ and\ \citenamefont {Mazziotti}(2021)}]{Smart2021}%
  \BibitemOpen
  \bibfield  {author} {\bibinfo {author} {\bibfnamefont {S.~E.}\ \bibnamefont {Smart}}\ and\ \bibinfo {author} {\bibfnamefont {D.~A.}\ \bibnamefont {Mazziotti}},\ }\bibfield  {title} {\bibinfo {title} {{Quantum Solver of Contracted Eigenvalue Equations for Scalable Molecular Simulations on Quantum Computing Devices}},\ }\href {https://doi.org/10.1103/PhysRevLett.126.070504} {\bibfield  {journal} {\bibinfo  {journal} {Phys. Rev. Lett.}\ }\textbf {\bibinfo {volume} {126}},\ \bibinfo {pages} {070504} (\bibinfo {year} {2021})}\BibitemShut {NoStop}%
\bibitem [{\citenamefont {Garrett}\ \emph {et~al.}(2026)\citenamefont {Garrett}, \citenamefont {Rose},\ and\ \citenamefont {Mazziotti}}]{Garrett2026}%
  \BibitemOpen
  \bibfield  {author} {\bibinfo {author} {\bibfnamefont {N.}~\bibnamefont {Garrett}}, \bibinfo {author} {\bibfnamefont {M.}~\bibnamefont {Rose}},\ and\ \bibinfo {author} {\bibfnamefont {D.~A.}\ \bibnamefont {Mazziotti}},\ }\bibfield  {title} {\bibinfo {title} {{Size-Consistent Quantum Chemistry on Quantum Computers}},\ } {\bibfield  {journal} {\bibinfo  {journal} {J. Phys. Chem. Lett.}\ }\textbf {\bibinfo {volume} {17}},\ \bibinfo {pages} {1892} (\bibinfo {year} {2026})},\ \Eprint {https://arxiv.org/abs/2512.18395} {2512.18395} \BibitemShut {NoStop}%
\bibitem [{\citenamefont {Tilly}\ \emph {et~al.}(2021)\citenamefont {Tilly}, \citenamefont {Sriluckshmy}, \citenamefont {Patel}, \citenamefont {Fontana}, \citenamefont {Rungger}, \citenamefont {Grant}, \citenamefont {Anderson}, \citenamefont {Tennyson},\ and\ \citenamefont {Booth}}]{Tilly2021}%
  \BibitemOpen
  \bibfield  {author} {\bibinfo {author} {\bibfnamefont {J.}~\bibnamefont {Tilly}}, \bibinfo {author} {\bibfnamefont {P.~V.}\ \bibnamefont {Sriluckshmy}}, \bibinfo {author} {\bibfnamefont {A.}~\bibnamefont {Patel}}, \bibinfo {author} {\bibfnamefont {E.}~\bibnamefont {Fontana}}, \bibinfo {author} {\bibfnamefont {I.}~\bibnamefont {Rungger}}, \bibinfo {author} {\bibfnamefont {E.}~\bibnamefont {Grant}}, \bibinfo {author} {\bibfnamefont {R.}~\bibnamefont {Anderson}}, \bibinfo {author} {\bibfnamefont {J.}~\bibnamefont {Tennyson}},\ and\ \bibinfo {author} {\bibfnamefont {G.~H.}\ \bibnamefont {Booth}},\ }\bibfield  {title} {\bibinfo {title} {{Reduced density matrix sampling: Self-consistent embedding and multiscale electronic structure on current generation quantum computers}},\ }\href {https://doi.org/10.1103/PhysRevResearch.3.033230} {\bibfield  {journal} {\bibinfo  {journal} {Phys. Rev. Res.}\ }\textbf {\bibinfo {volume} {3}},\ \bibinfo {pages} {1} (\bibinfo {year} {2021})}\BibitemShut {NoStop}%
\bibitem [{\citenamefont {Gilbert}(1975)}]{Gilbert1975}%
  \BibitemOpen
  \bibfield  {author} {\bibinfo {author} {\bibfnamefont {T.~L.}\ \bibnamefont {Gilbert}},\ }\bibfield  {title} {\bibinfo {title} {{Hohenberg-Kohn theorem for nonlocal external potentials}},\ }\href {https://doi.org/10.1103/PhysRevB.12.2111} {\bibfield  {journal} {\bibinfo  {journal} {Phys. Rev. B}\ }\textbf {\bibinfo {volume} {12}},\ \bibinfo {pages} {2111} (\bibinfo {year} {1975})}\BibitemShut {NoStop}%
\bibitem [{\citenamefont {Valone}(1980)}]{Valone1980}%
  \BibitemOpen
  \bibfield  {author} {\bibinfo {author} {\bibfnamefont {S.~M.}\ \bibnamefont {Valone}},\ }\bibfield  {title} {\bibinfo {title} {{Consequences of extending 1 matrix energy functionals pure-state representable to all ensemble representable 1 matrices}},\ }\href {https://doi.org/10.1063/1.440249} {\bibfield  {journal} {\bibinfo  {journal} {J. Chem. Phys.}\ }\textbf {\bibinfo {volume} {73}},\ \bibinfo {pages} {1344} (\bibinfo {year} {1980})}\BibitemShut {NoStop}%
\bibitem [{\citenamefont {Piris}(2024{\natexlab{a}})}]{Piris2024}%
  \BibitemOpen
  \bibfield  {author} {\bibinfo {author} {\bibfnamefont {M.}~\bibnamefont {Piris}},\ }\bibfield  {title} {\bibinfo {title} {{Advances in Approximate Natural Orbital Functionals: From Historical Perspectives to Contemporary Developments}},\ }\href {https://doi.org/10.1016/bs.aiq.2024.04.002} {\bibfield  {journal} {\bibinfo  {journal} {Adv. Quantum Chem.}\ }\textbf {\bibinfo {volume} {90}},\ \bibinfo {pages} {15} (\bibinfo {year} {2024}{\natexlab{a}})}\BibitemShut {NoStop}%
\bibitem [{\citenamefont {Piris}(2024{\natexlab{b}})}]{Piris2024a}%
  \BibitemOpen
  \bibfield  {author} {\bibinfo {author} {\bibfnamefont {M.}~\bibnamefont {Piris}},\ }\bibfield  {title} {\bibinfo {title} {Exploring the potential of natural orbital functionals},\ }\href {https://doi.org/10.1039/d4sc05810k} {\bibfield  {journal} {\bibinfo  {journal} {Chem. Sci.}\ }\textbf {\bibinfo {volume} {15}},\ \bibinfo {pages} {17284} (\bibinfo {year} {2024}{\natexlab{b}})}\BibitemShut {NoStop}%
\bibitem [{\citenamefont {Lew-Yee}\ \emph {et~al.}(2026)\citenamefont {Lew-Yee}, \citenamefont {Mitxelena}, \citenamefont {del Campo},\ and\ \citenamefont {Piris}}]{Lew-Yee2026}%
  \BibitemOpen
  \bibfield  {author} {\bibinfo {author} {\bibfnamefont {J.~F.~H.}\ \bibnamefont {Lew-Yee}}, \bibinfo {author} {\bibfnamefont {I.}~\bibnamefont {Mitxelena}}, \bibinfo {author} {\bibfnamefont {J.~M.}\ \bibnamefont {del Campo}},\ and\ \bibinfo {author} {\bibfnamefont {M.}~\bibnamefont {Piris}},\ }\bibfield  {title} {\bibinfo {title} {{DoNOF 2.0: A modern open-source electronic structure program for natural orbital functionals}},\ }\href {https://doi.org/10.1063/5.0316927} {\bibfield  {journal} {\bibinfo  {journal} {J. Chem. Phys.}\ }\textbf {\bibinfo {volume} {164}},\ \bibinfo {pages} {072501} (\bibinfo {year} {2026})}\BibitemShut {NoStop}%
\bibitem [{\citenamefont {Piris}\ \emph {et~al.}(2011)\citenamefont {Piris}, \citenamefont {Lopez}, \citenamefont {Ruip{\'{e}}rez}, \citenamefont {Matxain},\ and\ \citenamefont {Ugalde}}]{Piris2011}%
  \BibitemOpen
  \bibfield  {author} {\bibinfo {author} {\bibfnamefont {M.}~\bibnamefont {Piris}}, \bibinfo {author} {\bibfnamefont {X.}~\bibnamefont {Lopez}}, \bibinfo {author} {\bibfnamefont {F.}~\bibnamefont {Ruip{\'{e}}rez}}, \bibinfo {author} {\bibfnamefont {J.~M.}\ \bibnamefont {Matxain}},\ and\ \bibinfo {author} {\bibfnamefont {J.~M.}\ \bibnamefont {Ugalde}},\ }\bibfield  {title} {\bibinfo {title} {{A natural orbital functional for multiconfigurational states.}},\ }\href {https://doi.org/10.1063/1.3582792} {\bibfield  {journal} {\bibinfo  {journal} {J. Chem. Phys.}\ }\textbf {\bibinfo {volume} {134}},\ \bibinfo {pages} {164102} (\bibinfo {year} {2011})}\BibitemShut {NoStop}%
\bibitem [{\citenamefont {Piris}(2017)}]{Piris2017}%
  \BibitemOpen
  \bibfield  {author} {\bibinfo {author} {\bibfnamefont {M.}~\bibnamefont {Piris}},\ }\bibfield  {title} {\bibinfo {title} {{Global Method for Electron Correlation}},\ }\href {https://doi.org/10.1103/PhysRevLett.119.063002} {\bibfield  {journal} {\bibinfo  {journal} {Phys. Rev. Lett.}\ }\textbf {\bibinfo {volume} {119}},\ \bibinfo {pages} {063002} (\bibinfo {year} {2017})}\BibitemShut {NoStop}%
\bibitem [{\citenamefont {Piris}(2021)}]{Piris2021}%
  \BibitemOpen
  \bibfield  {author} {\bibinfo {author} {\bibfnamefont {M.}~\bibnamefont {Piris}},\ }\bibfield  {title} {\bibinfo {title} {{Global Natural Orbital Functional: Towards the Complete Description of the Electron Correlation}},\ }\href {https://doi.org/10.1103/PhysRevLett.127.233001} {\bibfield  {journal} {\bibinfo  {journal} {Phys. Rev. Lett.}\ }\textbf {\bibinfo {volume} {127}},\ \bibinfo {pages} {233001} (\bibinfo {year} {2021})}\BibitemShut {NoStop}%
\bibitem [{\citenamefont {Lew-Yee}\ \emph {et~al.}(2023{\natexlab{a}})\citenamefont {Lew-Yee}, \citenamefont {Piris},\ and\ \citenamefont {M.~del Campo}}]{Lew-Yee2023}%
  \BibitemOpen
  \bibfield  {author} {\bibinfo {author} {\bibfnamefont {J.~F.~H.}\ \bibnamefont {Lew-Yee}}, \bibinfo {author} {\bibfnamefont {M.}~\bibnamefont {Piris}},\ and\ \bibinfo {author} {\bibfnamefont {J.}~\bibnamefont {M.~del Campo}},\ }\bibfield  {title} {\bibinfo {title} {{Outstanding improvement in removing the delocalization error by global natural orbital functional}},\ }\href {https://doi.org/10.1063/5.0137378} {\bibfield  {journal} {\bibinfo  {journal} {J. Chem. Phys.}\ }\textbf {\bibinfo {volume} {158}},\ \bibinfo {pages} {084110} (\bibinfo {year} {2023}{\natexlab{a}})}\BibitemShut {NoStop}%
\bibitem [{\citenamefont {Lew-Yee}\ \emph {et~al.}(2024)\citenamefont {Lew-Yee}, \citenamefont {Bonfil-Rivera}, \citenamefont {Piris},\ and\ \citenamefont {del Campo}}]{Lew-Yee2024}%
  \BibitemOpen
  \bibfield  {author} {\bibinfo {author} {\bibfnamefont {J.~F.~H.}\ \bibnamefont {Lew-Yee}}, \bibinfo {author} {\bibfnamefont {I.~A.}\ \bibnamefont {Bonfil-Rivera}}, \bibinfo {author} {\bibfnamefont {M.}~\bibnamefont {Piris}},\ and\ \bibinfo {author} {\bibfnamefont {J.~M.}\ \bibnamefont {del Campo}},\ }\bibfield  {title} {\bibinfo {title} {{Excited States by Coupling Piris Natural Orbital Functionals with the Extended Random-Phase Approximation}},\ }\href {https://doi.org/10.1021/acs.jctc.3c01194} {\bibfield  {journal} {\bibinfo  {journal} {J. Chem. Theory Comput.}\ }\textbf {\bibinfo {volume} {20}},\ \bibinfo {pages} {2140} (\bibinfo {year} {2024})}\BibitemShut {NoStop}%
\bibitem [{\citenamefont {Mitxelena}\ \emph {et~al.}(2026)\citenamefont {Mitxelena}, \citenamefont {Lew-Yee},\ and\ \citenamefont {Piris}}]{Mitxelena2026}%
  \BibitemOpen
  \bibfield  {author} {\bibinfo {author} {\bibfnamefont {I.}~\bibnamefont {Mitxelena}}, \bibinfo {author} {\bibfnamefont {J.~F.~H.}\ \bibnamefont {Lew-Yee}},\ and\ \bibinfo {author} {\bibfnamefont {M.}~\bibnamefont {Piris}},\ }\bibfield  {title} {\bibinfo {title} {{5- and 6-membered rings: A natural orbital functional study}},\ }\href {https://doi.org/10.1021/acs.jctc.5c01861} {\bibfield  {journal} {\bibinfo  {journal} {J. Chem. Theory Comp.}\ }\textbf {\bibinfo {volume} {22}},\ \bibinfo {pages} {2799} (\bibinfo {year} {2026})}\BibitemShut {NoStop}%
\bibitem [{\citenamefont {Mitxelena}\ and\ \citenamefont {Piris}(2020{\natexlab{a}})}]{Mitxelena2020a}%
  \BibitemOpen
  \bibfield  {author} {\bibinfo {author} {\bibfnamefont {I.}~\bibnamefont {Mitxelena}}\ and\ \bibinfo {author} {\bibfnamefont {M.}~\bibnamefont {Piris}},\ }\bibfield  {title} {\bibinfo {title} {{An efficient method for strongly correlated electrons in one dimension}},\ }\href {https://doi.org/10.1088/1361-648X/ab6d11} {\bibfield  {journal} {\bibinfo  {journal} {J. Phys. Condens. Matter}\ }\textbf {\bibinfo {volume} {32}},\ \bibinfo {pages} {17LT01} (\bibinfo {year} {2020}{\natexlab{a}})}\BibitemShut {NoStop}%
\bibitem [{\citenamefont {Mitxelena}\ and\ \citenamefont {Piris}(2020{\natexlab{b}})}]{Mitxelena2020b}%
  \BibitemOpen
  \bibfield  {author} {\bibinfo {author} {\bibfnamefont {I.}~\bibnamefont {Mitxelena}}\ and\ \bibinfo {author} {\bibfnamefont {M.}~\bibnamefont {Piris}},\ }\bibfield  {title} {\bibinfo {title} {{An efficient method for strongly correlated electrons in two-dimensions}},\ }\href {https://doi.org/10.1063/1.5140985} {\bibfield  {journal} {\bibinfo  {journal} {J. Chem. Phys.}\ }\textbf {\bibinfo {volume} {152}},\ \bibinfo {pages} {064108} (\bibinfo {year} {2020}{\natexlab{b}})}\BibitemShut {NoStop}%
\bibitem [{\citenamefont {Mitxelena}\ and\ \citenamefont {Piris}(2022)}]{Mitxelena2022}%
  \BibitemOpen
  \bibfield  {author} {\bibinfo {author} {\bibfnamefont {I.}~\bibnamefont {Mitxelena}}\ and\ \bibinfo {author} {\bibfnamefont {M.}~\bibnamefont {Piris}},\ }\bibfield  {title} {\bibinfo {title} {{Benchmarking GNOF against FCI in challenging systems in one, two, and three dimensions}},\ }\href {https://doi.org/10.1063/5.0092611} {\bibfield  {journal} {\bibinfo  {journal} {J. Chem. Phys.}\ }\textbf {\bibinfo {volume} {156}},\ \bibinfo {pages} {214102} (\bibinfo {year} {2022})}\BibitemShut {NoStop}%
\bibitem [{\citenamefont {Mitxelena}\ and\ \citenamefont {Piris}(2024)}]{Mitxelena2024}%
  \BibitemOpen
  \bibfield  {author} {\bibinfo {author} {\bibfnamefont {I.}~\bibnamefont {Mitxelena}}\ and\ \bibinfo {author} {\bibfnamefont {M.}~\bibnamefont {Piris}},\ }\bibfield  {title} {\bibinfo {title} {Assessing the global natural orbital functional approximation on model systems with strong correlation},\ }\href {https://doi.org/10.1063/5.0207325} {\bibfield  {journal} {\bibinfo  {journal} {J. Chem. Phys.}\ }\textbf {\bibinfo {volume} {160}},\ \bibinfo {pages} {204106} (\bibinfo {year} {2024})}\BibitemShut {NoStop}%
\bibitem [{\citenamefont {Lew-Yee}\ \emph {et~al.}(2025)\citenamefont {Lew-Yee}, \citenamefont {M.~del Campo},\ and\ \citenamefont {Piris}}]{Lew-Yee2025b}%
  \BibitemOpen
  \bibfield  {author} {\bibinfo {author} {\bibfnamefont {J.~F.~H.}\ \bibnamefont {Lew-Yee}}, \bibinfo {author} {\bibfnamefont {J.}~\bibnamefont {M.~del Campo}},\ and\ \bibinfo {author} {\bibfnamefont {M.}~\bibnamefont {Piris}},\ }\bibfield  {title} {\bibinfo {title} {Advancing natural orbital functional calculations through deep learning-inspired techniques for large-scale strongly correlated electron systems},\ }\href {https://doi.org/10.1103/PhysRevLett.134.206401} {\bibfield  {journal} {\bibinfo  {journal} {Phys. Rev. Lett.}\ }\textbf {\bibinfo {volume} {134}},\ \bibinfo {pages} {206401} (\bibinfo {year} {2025})}\BibitemShut {NoStop}%
\bibitem [{\citenamefont {Lew-Yee}\ and\ \citenamefont {Piris}(2025{\natexlab{a}})}]{Lew-Yee2025c}%
  \BibitemOpen
  \bibfield  {author} {\bibinfo {author} {\bibfnamefont {J.~F.~H.}\ \bibnamefont {Lew-Yee}}\ and\ \bibinfo {author} {\bibfnamefont {M.}~\bibnamefont {Piris}},\ }\bibfield  {title} {\bibinfo {title} {Metal-insulator transition described by natural orbital functional theory},\ } {\bibfield  {journal} {\bibinfo  {journal} {Rev. Cubana de Fis.}\ }\textbf {\bibinfo {volume} {42}},\ \bibinfo {pages} {12} (\bibinfo {year} {2025}{\natexlab{a}})}\BibitemShut {NoStop}%
\bibitem [{\citenamefont {Motta}\ \emph {et~al.}(2024)\citenamefont {Motta}, \citenamefont {Kirby}, \citenamefont {Liepuoniute}, \citenamefont {Sung}, \citenamefont {Cohn}, \citenamefont {Mezzacapo}, \citenamefont {Klymko}, \citenamefont {Nguyen}, \citenamefont {Yoshioka},\ and\ \citenamefont {Rice}}]{Motta2024}%
  \BibitemOpen
  \bibfield  {author} {\bibinfo {author} {\bibfnamefont {M.}~\bibnamefont {Motta}}, \bibinfo {author} {\bibfnamefont {W.}~\bibnamefont {Kirby}}, \bibinfo {author} {\bibfnamefont {I.}~\bibnamefont {Liepuoniute}}, \bibinfo {author} {\bibfnamefont {K.~J.}\ \bibnamefont {Sung}}, \bibinfo {author} {\bibfnamefont {J.}~\bibnamefont {Cohn}}, \bibinfo {author} {\bibfnamefont {A.}~\bibnamefont {Mezzacapo}}, \bibinfo {author} {\bibfnamefont {K.}~\bibnamefont {Klymko}}, \bibinfo {author} {\bibfnamefont {N.}~\bibnamefont {Nguyen}}, \bibinfo {author} {\bibfnamefont {N.}~\bibnamefont {Yoshioka}},\ and\ \bibinfo {author} {\bibfnamefont {J.~E.}\ \bibnamefont {Rice}},\ }\bibfield  {title} {\bibinfo {title} {{Subspace methods for electronic structure simulations on quantum computers}},\ }\href {https://doi.org/10.1088/2516-1075/ad3592} {\bibfield  {journal} {\bibinfo  {journal} {Electron. Struct.}\ }\textbf {\bibinfo {volume} {6}},\ \bibinfo {pages} {013001} (\bibinfo {year} {2024})}\BibitemShut {NoStop}%
\bibitem [{\citenamefont {Patel}\ \emph {et~al.}(2026)\citenamefont {Patel}, \citenamefont {Jayakumar}, \citenamefont {Huang}, \citenamefont {Zeng},\ and\ \citenamefont {Izmaylov}}]{Patel2026}%
  \BibitemOpen
  \bibfield  {author} {\bibinfo {author} {\bibfnamefont {S.}~\bibnamefont {Patel}}, \bibinfo {author} {\bibfnamefont {P.}~\bibnamefont {Jayakumar}}, \bibinfo {author} {\bibfnamefont {R.}~\bibnamefont {Huang}}, \bibinfo {author} {\bibfnamefont {T.}~\bibnamefont {Zeng}},\ and\ \bibinfo {author} {\bibfnamefont {A.~F.}\ \bibnamefont {Izmaylov}},\ }\bibfield  {title} {\bibinfo {title} {{Quantum Seniority-Based Subspace Expansion: Linear Combinations of Short-Circuit Unitary Transformations for the Electronic Structure Problem}},\ }\href {https://doi.org/10.1021/acs.jctc.6c00017} {\bibfield  {journal} {\bibinfo  {journal} {J. Chem. Theory Comput.}\ }\textbf {\bibinfo {volume} {22}},\ \bibinfo {pages} {3937} (\bibinfo {year} {2026})}\BibitemShut {NoStop}%
\bibitem [{\citenamefont {Lew-Yee}\ and\ \citenamefont {Piris}(2025{\natexlab{b}})}]{Lew-Yee2025}%
  \BibitemOpen
  \bibfield  {author} {\bibinfo {author} {\bibfnamefont {J.~F.~H.}\ \bibnamefont {Lew-Yee}}\ and\ \bibinfo {author} {\bibfnamefont {M.}~\bibnamefont {Piris}},\ }\bibfield  {title} {\bibinfo {title} {{Efficient Energy Measurement of Chemical Systems via One-Particle Reduced Density Matrix: A NOF-VQE Approach for Optimized Sampling}},\ }\href {https://doi.org/10.1021/acs.jctc.4c01734} {\bibfield  {journal} {\bibinfo  {journal} {J. Chem. Theory Comp.}\ }\textbf {\bibinfo {volume} {21}},\ \bibinfo {pages} {2402} (\bibinfo {year} {2025}{\natexlab{b}})}\BibitemShut {NoStop}%
\bibitem [{\citenamefont {Piris}(2018{\natexlab{a}})}]{Piris2018e}%
  \BibitemOpen
  \bibfield  {author} {\bibinfo {author} {\bibfnamefont {M.}~\bibnamefont {Piris}},\ }\bibfield  {title} {\bibinfo {title} {{The electron pairing approach in NOF Theory}},\ }in\ {\emph {\bibinfo {booktitle} {Quantum Chemistry at the Dawn of the 21st Century. Series: Innovations in Computational Chemistry}}},\ \bibinfo {editor} {edited by\ \bibinfo {editor} {\bibfnamefont {R.}~\bibnamefont {Carb{\'{o}}-Dorca}}\ and\ \bibinfo {editor} {\bibfnamefont {T.}~\bibnamefont {Chakraborty}}}\ (\bibinfo  {publisher} {Apple Academic Press},\ \bibinfo {year} {2018})\ Chap.~\bibinfo {chapter} {22}, pp.\ \bibinfo {pages} {593--620}\BibitemShut {NoStop}%
\bibitem [{\citenamefont {Piris}\ \emph {et~al.}(2013)\citenamefont {Piris}, \citenamefont {Matxain},\ and\ \citenamefont {Lopez}}]{Piris2013a}%
  \BibitemOpen
  \bibfield  {author} {\bibinfo {author} {\bibfnamefont {M.}~\bibnamefont {Piris}}, \bibinfo {author} {\bibfnamefont {J.~M.}\ \bibnamefont {Matxain}},\ and\ \bibinfo {author} {\bibfnamefont {X.}~\bibnamefont {Lopez}},\ }\bibfield  {title} {\bibinfo {title} {The intrapair electron correlation in natural orbital functional theory},\ }\href {https://doi.org/10.1063/1.4844075} {\bibfield  {journal} {\bibinfo  {journal} {J. Chem. Phys.}\ }\textbf {\bibinfo {volume} {139}},\ \bibinfo {pages} {234109} (\bibinfo {year} {2013})}\BibitemShut {NoStop}%
\bibitem [{\citenamefont {Piris}(2013)}]{Piris2013c}%
  \BibitemOpen
  \bibfield  {author} {\bibinfo {author} {\bibfnamefont {M.}~\bibnamefont {Piris}},\ }\bibfield  {title} {\bibinfo {title} {{Interpair electron correlation by second-order perturbative corrections to PNOF5}},\ }\href {https://doi.org/10.1063/1.4817946} {\bibfield  {journal} {\bibinfo  {journal} {J. Chem. Phys.}\ }\textbf {\bibinfo {volume} {139}},\ \bibinfo {pages} {064111} (\bibinfo {year} {2013})}\BibitemShut {NoStop}%
\bibitem [{\citenamefont {Mitxelena}\ \emph {et~al.}(2018)\citenamefont {Mitxelena}, \citenamefont {Rodr{\'{i}}guez-Mayorga},\ and\ \citenamefont {Piris}}]{Mitxelena2018a}%
  \BibitemOpen
  \bibfield  {author} {\bibinfo {author} {\bibfnamefont {I.}~\bibnamefont {Mitxelena}}, \bibinfo {author} {\bibfnamefont {M.}~\bibnamefont {Rodr{\'{i}}guez-Mayorga}},\ and\ \bibinfo {author} {\bibfnamefont {M.}~\bibnamefont {Piris}},\ }\bibfield  {title} {\bibinfo {title} {{Phase Dilemma in Natural Orbital Functional Theory from the N-representability Perspective}},\ }\href {https://doi.org/10.1140/epjb/e2018-90078-8} {\bibfield  {journal} {\bibinfo  {journal} {Eur. Phys. J. B}\ }\textbf {\bibinfo {volume} {91}},\ \bibinfo {pages} {109} (\bibinfo {year} {2018})}\BibitemShut {NoStop}%
\bibitem [{\citenamefont {Piris}(2018{\natexlab{b}})}]{Piris2018b}%
  \BibitemOpen
  \bibfield  {author} {\bibinfo {author} {\bibfnamefont {M.}~\bibnamefont {Piris}},\ }\bibfield  {title} {\bibinfo {title} {{Dynamic electron-correlation energy in the natural-orbital-functional second-order-M{\o}ller-Plesset method from the orbital-invariant perturbation theory}},\ }\href {https://doi.org/10.1103/PhysRevA.98.022504} {\bibfield  {journal} {\bibinfo  {journal} {Phys. Rev. A}\ }\textbf {\bibinfo {volume} {98}},\ \bibinfo {pages} {022504} (\bibinfo {year} {2018}{\natexlab{b}})}\BibitemShut {NoStop}%
\bibitem [{\citenamefont {Lew-Yee}\ \emph {et~al.}(2023{\natexlab{b}})\citenamefont {Lew-Yee}, \citenamefont {M.~del Campo},\ and\ \citenamefont {Piris}}]{Lew-Yee2023a}%
  \BibitemOpen
  \bibfield  {author} {\bibinfo {author} {\bibfnamefont {J.~F.~H.}\ \bibnamefont {Lew-Yee}}, \bibinfo {author} {\bibfnamefont {J.}~\bibnamefont {M.~del Campo}},\ and\ \bibinfo {author} {\bibfnamefont {M.}~\bibnamefont {Piris}},\ }\bibfield  {title} {\bibinfo {title} {{Electron Correlation in the Iron(II) Porphyrin by Natural Orbital Functional Approximations}},\ }\href {https://doi.org/10.1021/acs.jctc.2c01093} {\bibfield  {journal} {\bibinfo  {journal} {J. Chem. Theory Comput.}\ }\textbf {\bibinfo {volume} {19}},\ \bibinfo {pages} {211} (\bibinfo {year} {2023}{\natexlab{b}})}\BibitemShut {NoStop}%
\bibitem [{\citenamefont {Rivero-Santamaría}\ and\ \citenamefont {Piris}(2024)}]{RiveroSantamaria2024}%
  \BibitemOpen
  \bibfield  {author} {\bibinfo {author} {\bibfnamefont {A.}~\bibnamefont {Rivero-Santamaría}}\ and\ \bibinfo {author} {\bibfnamefont {M.}~\bibnamefont {Piris}},\ }\bibfield  {title} {\bibinfo {title} {Time evolution of natural orbitals in ab initio molecular dynamics},\ }\href {https://doi.org/10.1063/5.0188491} {\bibfield  {journal} {\bibinfo  {journal} {J. Chem. Phys.}\ }\textbf {\bibinfo {volume} {160}},\ \bibinfo {pages} {071102} (\bibinfo {year} {2024})}\BibitemShut {NoStop}%
\bibitem [{\citenamefont {Piris}\ \emph {et~al.}(2024)\citenamefont {Piris}, \citenamefont {Lopez},\ and\ \citenamefont {Ugalde}}]{Piris2024b}%
  \BibitemOpen
  \bibfield  {author} {\bibinfo {author} {\bibfnamefont {M.}~\bibnamefont {Piris}}, \bibinfo {author} {\bibfnamefont {X.}~\bibnamefont {Lopez}},\ and\ \bibinfo {author} {\bibfnamefont {J.~M.}\ \bibnamefont {Ugalde}},\ }\bibfield  {title} {\bibinfo {title} {Time-resolved chemical bonding structure evolution by direct-dynamics chemical simulations},\ }\href {https://doi.org/10.1021/acs.jpclett.4c03010} {\bibfield  {journal} {\bibinfo  {journal} {J. Phys. Chem. Lett.}\ }\textbf {\bibinfo {volume} {15}},\ \bibinfo {pages} {12138} (\bibinfo {year} {2024})}\BibitemShut {NoStop}%
\bibitem [{\citenamefont {Piris}(2019)}]{Piris2019}%
  \BibitemOpen
  \bibfield  {author} {\bibinfo {author} {\bibfnamefont {M.}~\bibnamefont {Piris}},\ }\bibfield  {title} {\bibinfo {title} {{Natural orbital functional for multiplets}},\ }\href {https://doi.org/10.1103/PhysRevA.100.032508} {\bibfield  {journal} {\bibinfo  {journal} {Phys. Rev. A}\ }\textbf {\bibinfo {volume} {100}},\ \bibinfo {pages} {32508} (\bibinfo {year} {2019})}\BibitemShut {NoStop}%
\bibitem [{\citenamefont {Surjan}(1999)}]{Surjan1999}%
  \BibitemOpen
  \bibfield  {author} {\bibinfo {author} {\bibfnamefont {P.~R.}\ \bibnamefont {Surjan}},\ }\bibfield  {title} {\bibinfo {title} {{An Introduction to the Theory of Geminals}},\ }in\ {\emph {\bibinfo {booktitle} {Topics in Current Chemistry, Vol. 203}}}\ (\bibinfo  {publisher} {Springer-Verlag Berlin Heidelberg},\ \bibinfo {year} {1999})\ pp.\ \bibinfo {pages} {63--88}\BibitemShut {NoStop}%
\bibitem [{\citenamefont {Salazar}\ \emph {et~al.}(2026)\citenamefont {Salazar}, \citenamefont {Huan Lew-Yee},\ and\ \citenamefont {Piris}}]{Salazar2026ONVQE}%
  \BibitemOpen
  \bibfield  {author} {\bibinfo {author} {\bibfnamefont {E.~X.}\ \bibnamefont {Salazar}}, \bibinfo {author} {\bibfnamefont {J.~F.}\ \bibnamefont {Huan Lew-Yee}},\ and\ \bibinfo {author} {\bibfnamefont {M.}~\bibnamefont {Piris}},\ } {\bibinfo {title} {{ON-VQE}: Occupation Number Variational Quantum Eigensolver}},\ \bibinfo {howpublished} {\url{https://github.com/gatox/PennyLane_Exercises/blob/main/test_nof_vqe/NOFVQE/nofvqe.py}} (\bibinfo {year} {2026}),\ \bibinfo {note} {open-source implementation}\BibitemShut {NoStop}%
\bibitem [{\citenamefont {Asadi}\ \emph {et~al.}(2024)\citenamefont {Asadi}, \citenamefont {Dusko}, \citenamefont {Park}, \citenamefont {Michaud-Rioux}, \citenamefont {Schoch}, \citenamefont {Shu}, \citenamefont {Vincent},\ and\ \citenamefont {O'Riordan}}]{Asadi2024}%
  \BibitemOpen
  \bibfield  {author} {\bibinfo {author} {\bibfnamefont {A.}~\bibnamefont {Asadi}}, \bibinfo {author} {\bibfnamefont {A.}~\bibnamefont {Dusko}}, \bibinfo {author} {\bibfnamefont {C.-Y.}\ \bibnamefont {Park}}, \bibinfo {author} {\bibfnamefont {V.}~\bibnamefont {Michaud-Rioux}}, \bibinfo {author} {\bibfnamefont {I.}~\bibnamefont {Schoch}}, \bibinfo {author} {\bibfnamefont {S.}~\bibnamefont {Shu}}, \bibinfo {author} {\bibfnamefont {T.}~\bibnamefont {Vincent}},\ and\ \bibinfo {author} {\bibfnamefont {L.~J.}\ \bibnamefont {O'Riordan}},\ } {\bibinfo {title} {Hybrid quantum programming with pennylane lightning on hpc platforms}} (\bibinfo {year} {2024}),\ \Eprint {https://arxiv.org/abs/2403.02512} {arXiv:2403.02512 [quant-ph]} \BibitemShut {NoStop}%
\end{thebibliography}
\end{document}